\documentstyle{llncs}

\begin{document}
\input epsf

\title{The saturation threshold of public opinion: are aggressive media campaigns always effective?}

\author{Floriana Gargiulo\inst{1}, Stefano Lottini\inst{2}, Alberto Mazzoni\inst{1}}

\institute{Institute for Scientific Interchange, Torino, Italy \and
Universit\'{a} degli Studi di Torino, Italy \& INFN, Sez. di Torino}

\maketitle
\begin{abstract}
An important aspect that must be considered when studying opinion
formation phenomena is the different social attitude of the agents
taking part in the process. Different kinds of interconnections and
of interacting behaviours should be associated to the agents
depending on their opinion: radicals tend to self-segregate but, on
the other hand, have a stronger capacity to convince neutral agents.
Other important questions arise when the official media strongly
promotes one position. The different perception that each agent has
of the official information can lead, in case of monopolistic and
aggressive media, to a reaction effect in the population that starts
to create massive antagonist clusters.
\end{abstract}
\section{Introduction}
In 1969 Nixon introduced the concept of ``silent majority'':
Demonstrations against the war in Vietnam were taking place all
around the US and to enforce his warmonger decision Nixon exploited
the neutral, inactive position of the silent majority that was not
participating to the riots: those who do not protest support the
war. The wrong passage in this argument is the assumption that the
silent majority was supporting the war: when a state enters a war,
when politicians decide laws, when a company decides a shake-out,
when any group has to make a common decision, some people will line
up on one position, others will play an antagonist role but the
larger part of the people will not decide, simply adopting a neutral position.\\
The delegation mechanism induces people not to have their own
opinion on the single topics, to remain inactive in order to avoid
conflict and to move the discussion mechanism to higher political
levels. The radical position formation, instead, often requires a more or less
deep study and documentation phase. This different kind of choice,
neutrality vs.~radicalism, implies a different communicative
attitude between different kinds of agents: while radicals will
prefer to be surrounded by people sharing similar ideas, a neutral
agent will not have any \textit{a priori} preclusion in interacting
with anybody. On the other side, when a neutral agent interacts with
a radical one, the formation process underwent by the radical in the
decisional phase will give him bigger chance
to have a strong influence on neutral agents.\\
We consider these different communication attitudes and we show
that, in absence of external forcing, a pluralistic situation (where
many opinions persist) is reached only when the extremists are
integrated in the society and strongly consistent on their
positions.\\ In presence of media campaigns supporting one of the
positions, instead, we find that as the power of the media grows, a
secondary antagonist cluster appears to contrast the monopolistic
forcing, even in the setting that would have lead to total consensus
without media.

\section{Model description}
Many model have been constructed to describe opinion formation
processes in presence of extremist groups, such as, for example, the one in
\cite{deffuant_extr}. We want to consider this kind of analyses,
where different rules for social influence are implemented, in a
more complete framework where the structure of the social
network is also considered.\\
We consider a set of agents with opinion $o_i$ randomly extracted in
the range $[-1,1]$. We term ``radical'' an agent with opinion near 
the extremes, and ``neutral'' an agent with opinion approximatively
zero. We start from the assumption that a radical agent will
preferentially link to radicals, while neutrals will choose their
relationships in a completely random fashion.\footnote{To model this
situation, we start from a set of randomly connected nodes and we
add, one by one, new nodes with random opinion $o_N$ connecting the
new nodes to the pre--existing ones with probability: $P_{N\rightarrow
i} \propto \exp[-\beta \cdot o_N(o_N-o_i)]$.  We fix the parameter of the model to
$\beta=3$ and work on a system with $N=1000$ agents.}\\
As for the dynamical part, we refer again to \cite{noi}. A tolerance,
depending on the opinion, is associated to each agent:
\begin{equation}\label{toll}
    t_i=1-\alpha|o_i| \qquad  \alpha\in [0,1]\;\; .
\end{equation}
Two agents interact only if $|o_i-o_j|<\min(t_i,t_j)$. If the
agents interact, an asymmetric shift of the opinion in the radical
direction is performed:\footnote{In the following, the constant $\mu$ is always kept
fixed to its maximum value $0.5$, since apparently only the speed at which the 
final state is reached is sensitive to it, and not the final state itself.}
\begin{equation}\label{shift}
    o_i\rightarrow o_i+\mu \cdot t_i(o_j-o_i) \;\; .
\end{equation}
The parameter $\alpha$ tunes how the dynamics is sensitive to the
opinion difference: if $\alpha=0$ we have the standard Deffuant
model \cite{deffuant} (at its very critical point), characterised by 
uniform tolerance and symmetric drift after
the interaction. For $\alpha=1$ we have the maximum differentiation
of the behaviours of the agents: a radical will discuss only with
very similar agents and a more extreme position
will have a stronger attractive strength in the dynamics.\\
We observe a critical value of the dynamical parameter, $\alpha_c \simeq 0.8$,
such that for  $\alpha<\alpha_c$  the system converges to a single opinion
while, for $\alpha>\alpha_c$, the final state shows a larger number
of final-state opinions, as is illustrated in the leftmost 
plots of Fig.~\ref{fig:figure}.
This behaviour is due to the fact that the network structure is
strongly correlated to the initial opinions of the agents. Such
correlation permits the existence of gradual paths of communication
that always lead to convergence. The survival of minority clusters
for $\alpha>\alpha_c$  is due to the fact that, in this situation,
the most extremist agents are not involved in the global opinion
dynamics process since their tolerance is too small to interact with
someone far from their ideas;  they just interact with very
similar agents creating a sort of opinion niche. From this
reasoning, it follows also that the number of such sub--communities
increases with $\alpha$, while the population of each sub--community
decreases.
\subsection{The media influence}
\begin{center}\begin{figure}
    \epsfysize=7cm
   \centerline{\epsfbox{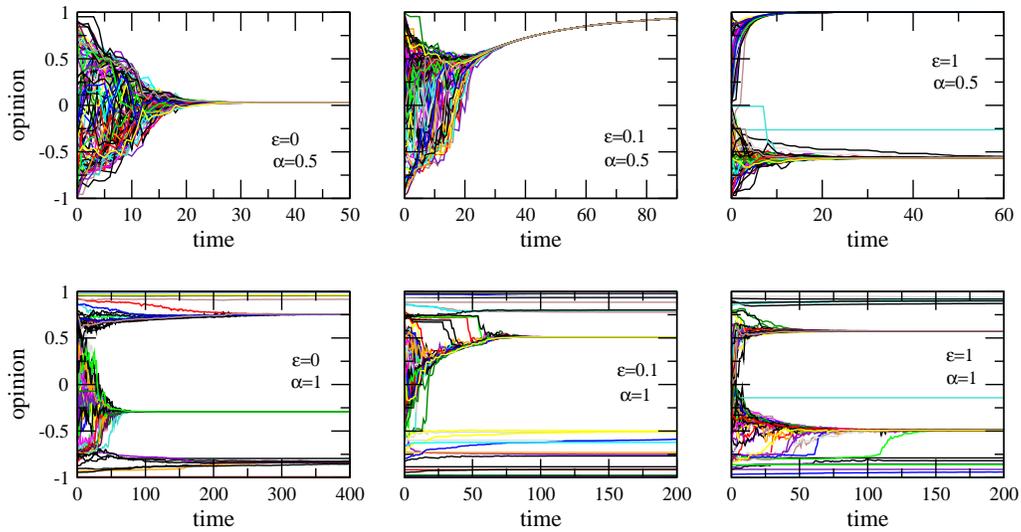}\hspace{0.5cm}}
        % altezza 3 cm
   \caption{\textit{(colour online)} Agents' opinions as a function of time. The upper plots refer to
        $\alpha=0.5$ and the lower plots to $\alpha=1$. From left to right
        we have: $\varepsilon= 0, 0.1, 1\;$.}
    \label{fig:figure}
 \end{figure}\end{center}

We want to investigate the effect on this model of a strong media
campaign supporting one of the two extremal positions.\\
Several mechanisms can explain the media influence on people: it can
be considered an indirect effect, like the so called \emph{third
person effect} \cite{3p}, where each agent feels the influence of the
media on the others and not on herself. Alternatively, it can be considered a
direct effect represented by some ``Big--Agent'' completely connected with the
whole network. We will consider only the latter mechanism introducing a
``Big--Agent'' (BA) with opinion $o_{BA}=+1$. This agent interacts with
all agents at every step of the opinion dynamics (namely, every
$n_{agent}$ interactions), with the usual tolerance rule
(\ref{toll}) and without changing her opinion at all. The strength
of the campaign (sensitive, for example, to whether it is also
supported by police repression of the opposite idea) affects the
interaction of the media with any agent: we model this persuasive
strength by introducing a parameter $\varepsilon\in[0,1]$ in the
opinion update rule (\ref{shift}):
\begin{equation}\label{shift2}
    o_i\rightarrow o_i+\mu \cdot \varepsilon t_i(o_{BA}-o_i) \;\; .
\end{equation}
For $\varepsilon=0$, the media pressure is null and we find the basic
model; as $\varepsilon$ grows, the media campaign becomes more
aggressive. \\
For any value of $\varepsilon>0$ the central cluster moves to the
BA's opinion, so that the neutral position is no longer
represented. The surprising result is that for $\varepsilon\rightarrow
1$ the size of the biggest cluster (that which supports the BA's
opinion) decreases, while a more and more populated antagonist cluster
appears near the opposite extreme (Fig.~\ref{fig:figure}).\\
This kind of result, where a media influence excessively strong does 
not lead to consensus, is in agreement with the result previously obtained in
\cite{MIAxel} for the case of Axelrod model.
%
% ---- Bibliography ----
%

\end{document}